\documentstyle[12pt]{article}
\begin{document}
\begin{flushright}
VAND-TH-96-4\\
LAEFF 96/22\\
September 1996\\
{\tt Revised December 1996}
\end{flushright}

\vspace{.5cm}
\begin{center}
{\large Representing Structural Information of
Helical Charge Distributions in Cylindrical
Coordinates }
\end{center}
\vspace{1cm}
\begin{center}
David Hochberg{$^{\dag,}$}\footnote{hochberg@laeff.esa.es},
Glenn Edwards{$^{*,}$}\footnote{edwardg1@ctrvax.vanderbilt.edu}
and
Thomas W. Kephart{$^{*,}$}\footnote{kephartt@ctrvax.vanderbilt.edu}\\
{$^\dag$} Laboratory for Space Astrophysics and Fundamental Physics, INTA\\
Apartado 50727, 28080 Madrid, Spain\\
{$^*$} Department of Physics and Astronomy, Vanderbilt University\\
Nashville, Tennessee 37235, USA
\end{center}

\begin{abstract}
Structural information in 
the local electric field produced by
helical charge distributions, such as dissolved DNA, is revealed
in a straightforward manner employing cylindrical coordinates. 
Comparison of structure factors derived in terms of cylindrical and
helical coordinates is made. A simple coordinate transformation serves
to relate the Green function in cylindrical and helical coordinates.
We also compare the electric field on the 
central axis of a single helix as calculated in both systems.
\end{abstract} 

\vfill\eject

\noindent

A few years ago we developed an exact analytical solution for 
a model of the local electric
potential and field arising from the double helix of phosphate groups
of a single B-DNA molecule immersed in an aqueous solvent \cite{HKE94}.
We subsequently extended our calculation to 
treat the full (sugar-phosphate plus base-pairs) discrete charge
distribution of homopolymer B-DNA in a solvent modelled by 
concentric dielectric cylinders \cite{EHK94}. 
In both cases we found a characteristic
length scale of $\approx 5 \AA$ associated with the radial persistence
length of either the helical imprint or
individual base-pair identity.
These detailed calculations are based on a theoretical
model of B-DNA in solution together with Green function techniques
to account for the contribution of each individual (partial) charge
to the full net potential. 
As we discussed in detail in \cite{HKE94,EHK94}, although
the helical configuration of point charges does not possess cylindrical
symmetry, the individual charges making up the backbone as well as the 
base-pairs can be assigned to a set of concentric cylindrical
surfaces and thus cylindrical coordinates provide the most natural 
system in which to {\it separate} Laplace's and Poisson's equations.
The full power of the Green function technique can be brought to bear
on the problem, which is made nontrivial by the presence of a non-uniform
dielectric. 
Moreover, in cylindrical coordinates one may easily identify
and separate out the featureless zero-mode contribution to the potential.
The zero mode, which goes as $\sim \ln(\rho)$,  
corresponds to the cylindrically symmetric
potential from a straight line of charge and is
what one sees far from the DNA surface. The higher-mode terms in the potential
therefore encode and reveal the specific helical conformation of the
molecule. The structural information contained in these higher modes is
complete, transparent and calculable \cite{HKE94,EHK94}. 

Though the above points have been adequately discussed in \cite{HKE94,EHK94},
we wish to emphasize them in light of a recent claim that 
structural information of (double) helix charge distributions is better
represented employing helical rather than cylindrical coordinates \cite{LCR}.
There, it was argued that structure factors derived in a certain helical
coordinate system reveal structural information in a 
more ``transparent'' 
fashion. The purpose of the present brief report is to demonstrate that 
similar structural information may be represented just as
easily in cylindrical coordinates.

\noindent
{\bf Structure Factors}

We turn to the derivation of structure factors in cylindrical
coordinates.
In the interests of clarity and brevity, we shall restrict 
our attention to the case
of a \underline{single} helix immersed 
in a \underline{single} dielectric background medium.
In our two previous papers 
we calculated explicitly both the electrostatic potential and 
electric field components
due to {\em many} helices 
(to account for the phosphate backbones and base pairs) immersed
in a medium modeled by a triply piecewise constant dielectric function.
However, since the single helix in a uniform background is the only case
treated by the methods in \cite{LCR},
we will stick to this simple example
to facilitate comparison.

Before tackling the full helix, 
we begin by considering the electrostatic potential for a {\em single} point
charge $q$ embedded in a uniform dielectric medium $\epsilon$, and
located at the point $(a,\phi',z')$. 
>From \cite{Jack}
or the expression (B26) in the Appendix of \cite{HKE94}, we have
\begin{equation}
\Phi(\rho,\phi,z) = \frac{4q}{\pi \epsilon}
{\sum^{\infty}_{m=0}}' \int^{\infty}_0 dk\, I_m(k\rho_{<})
K_m(k\rho_{>}) \cos(k[z-z'])) \cos(m[\phi - \phi']),
\end{equation}
where $I_m,K_m$ are the modified Bessel functions of integer order
$m$ and the prime on the sum indicates that the zero mode is to
be divided by two.
>From this we can build up the electrostatic potential corresponding to a
finite or an infinite distribution of (identical) point charges
distributed along a helix of constant radius $a$ and pitch $P$. 
We regard a single helix of evenly spaced point charges as
a two-dimensional regular lattice wrapped around the cylinder $\rho = a$.
Instead of summing the charges along the helix, we decompose 
this equivalent lattice
into a finite collection of one-dimensional vertical line charges that run
parallel to the $z$-axis and then sum over this ``bundle'' of one-dimensional
chains. Denote the number of such chains by the integer $N_o$ and the number
of charges along any such 
chain (this can be finite or infinite) by $2N+1$. The total
number of charges living on the helix is therefore 
equal to $N_o \times (2N+1)$. 
We now make this conceptual decomposition quantitative: the position 
in cylindrical coordinates of the
$n$-th charge on the $s$-th vertical line with fixed angular coordinate
$\phi_s$ is 
as follows: $(a, \phi_s, z_{n,s})$, where 
\begin{eqnarray}
\phi_s &=& (\frac{2\pi}{N_o})s = (\frac{2\pi \Delta z}{P}) s,\\ \nonumber
z_{n,s} &=& nP + s\Delta z,
\end{eqnarray}
for $0 \leq s \leq N_o -1$ and $-N \leq n \leq N$, 
where $\Delta z$ is the vertical rise per residue, and $N_o = P/{\Delta z}$
is the number of equally spaced residues per pitch length of the helix (see
Fig. 1 of ref. [1]). 
The single helix potential is obtained simply by replacing
$$ z' \longrightarrow z_{n,s}, \,\, {\rm and} \,\, 
\phi' \longrightarrow \phi_s$$
in (1) and summing over the point charges:
\begin{equation}
\Phi_{Helix}(\rho,\phi,z) = \sum^{N}_{n=-N} \sum^{N_o -1}_{s=0}
\Phi(\rho,\phi,z|a,\phi_s,z_{n,s}).
\end{equation}

This leads to the following expression for the electrostatic 
potential from a single
charged helix in a uniform background:
\begin{eqnarray}
\Phi_{Helix}(\rho,\phi,z) &=& \frac{2q}{\pi \epsilon}
{\sum^{\infty}_{m=0}}' \int^{\infty}_0 dk\, G_1(k) \sum_{\pm} \{ 
 C^{\pm}(m,k)\cos(kz \pm m\phi) \\ \nonumber 
&+& S^{\pm}(m,k)
\sin(kz \pm m\phi) \}
I_m(k\rho_{<})
K_m(k\rho_{>}).
\end{eqnarray}
The momentum dependent structure factors appearing here are defined as follows:
\begin{eqnarray}
G_1(k) &=& \frac{\sin [(2N+1)\frac{kP}{2}]}{\sin (\frac{kP}{2})}, \qquad
({\rm for}\, N \, {\rm finite}) \\ \nonumber
       &=& \frac{2\pi}{P} \sum^{\infty}_{j = -\infty} 
\delta(k - \frac{2\pi j}{P}),
\qquad ({\rm for}\, N \rightarrow \infty),
\end{eqnarray}
and,
\begin{eqnarray}
C^{\pm}(m,k) &=& G^{\pm}_2(m,k) \cos[(N_o -1)(k \pm \frac{2\pi m}{P})
\Delta z/2],\\ \nonumber
S^{\pm}(m,k) &=& G^{\pm}_2(m,k) \sin[(N_o -1)(k \pm \frac{2\pi m}{P})
\Delta z/2].
\end{eqnarray}
The remaining structure factor appearing in (4) is defined by
\begin{equation}
G_2^{\pm}(m,k) = \frac{\sin[ \frac{1}{2}(kP \pm 2\pi m)]}{\sin [\frac{1}{2N_o}
(kP \pm 2\pi m)]}. 
\end{equation}
The final form in which we have written (4) results from working out the double
sum in (3) and using one or more elementary series and/or trigonometric
identities, 
as well as the Poisson summation formula, as for example, in (A3) of
\cite{HKE94}.

By inspection of (4), it is clear that structural 
information (as defined by the
appearance of these so-called structure factors) of helical 
charge distributions
is revealed in the cylindrical coordinate
system. The structure factors depend explicitly on the pitch $P$ and the
rise per charge, $\Delta z$, which jointly 
parametrize the geometry of the helix. 

Note that our $G_1(k)$ has the same mathematical
form as the function $H_2(M)$ calculated in \cite{LCR}. 
The reason for this 
coincidence is clear: both structure
factors result from performing the charge sum (whether finite or not) 
over (half) the
{\em coordinate lines} corresponding to each 
coordinate system. That is, we
obtain $G_1$ from summing charges along the 
coordinate line $\theta = const.$ (a line parallel to
the $z$-axis) 
whereas 
$H_2$ results by summing along the coordinate 
line $t = const.$, that is, along a helix.
The linear momentum in the $\hat z$-direction is $k$, while that in the helical
direction $\hat t$ is $M$. One has in fact the formal correspondence between 
summing over vertical lines and summing over helices:
\begin{eqnarray}
k \leftrightarrow M, \\ \nonumber
P \leftrightarrow {\bar s}.
\end{eqnarray}

In the helical coordinate system, $H_2(M)$ is the full structure 
factor for a single
helix. In cylindrical coordinates, $G_1(k)$ gives the structure factor 
for a line charge, and
is not the complete ``answer'', as it were. That is why we need the 
additional factors
$C^{\pm}(m,k)$ and $S^{\pm}(m,k)$, as 
can be checked, by summing over the
$N_o$ vertical chains. In the cylindrical system, the helical 
structure is encoded
in terms of the factorized set of structure factors $G_1(k)C^{\pm}(m,k)$ and
$G_1(k)S^{\pm}(m,k)$. The necessity of the $C^{\pm},S^{\pm}$  
could have been anticipated on general grounds if one 
thinks of Fourier analysis on a cylinder, namely, one expands 
an arbitrary function of $\phi$ and $z$ in terms of the complete
set of functions of chiral ``up'' and ``down'' moving ``waves'':
$\{ \sin(kz \pm m\phi), \cos(kz \pm m\phi) \}$. 
These coefficient functions $(C^{\pm},S^{\pm})$ are simply the projections
of helical structure along the two helical directions, but expressed
in cylinder coordinates.
This is why these basis functions appear in our $\Phi_{Helix}$. We had
already made use of this fact during an intermediate stage of our published
calculation (see, e.g., Eq. (B8) in \cite{HKE94}). 

We may derive another form for the structure factor 
by performing the sum over charges in a
distinct way.
For example, if  we now sum the point charges {\em along}
the helix, then an equally transparent factor results--even using cylindrical
coordinates. Only a single sum need now be carried out because the 
helix sum means we are to make the replacements
\begin{eqnarray}
\phi' &\rightarrow& \phi_n = n \Delta \phi, \\ \nonumber
z' &\rightarrow&  z_n =  n \Delta z,
\end{eqnarray}
in (1)
for $-M \leq n \leq M$ (not be be confused with the 
momentum variable $M$ appearing in (8))
and sum over $n$ to obtain the corresponding
single helix potential. The origin charge corresponds to $n=0$ and has
coordinates $(a,0,0)$. Note that for a finite helix, the net charge
must turn out to be the same no matter how we carry out the sum. This
implies that $2M + 1 = N_o \times (2N + 1)$. By going through identical
steps as above, we find that the single helix potential is now represented
as
\begin{equation}
\Phi_{Helix}(\rho,\phi,z) = \frac{2q}{\pi \epsilon}
{\sum^{\infty}_{m=0}}' \int^{\infty}_0 dk\, \sum_{\pm} \{ 
 G_3^{\pm}(m,k)\cos(kz \pm m\phi) \}
I_m(k\rho_{<})
K_m(k\rho_{>}).
\end{equation}
Here, the structure factors turn out to be as follows:
\begin{equation}
G^{\pm}_3(m,k) = \frac{\sin[\frac{1}{2}(2M+1)(k\Delta z \pm m\Delta \phi)]}
{\sin[\frac{1}{2}(k\Delta z \pm m\Delta \phi)]},
\end{equation}
for finite $M$ (and hence, finite $N$) or,
\begin{equation}
G^{\pm}_3(m,k) = 2\pi \sum_{j= -\infty}^{\infty}
\delta (k\Delta z \pm m \Delta \phi -
2\pi j),
\end{equation}
for $M$ (or $N$) infinite. Again, the basic parameters $\Delta z,P$
describing the helix geometry are manifest in the cylinder-based 
structure factors.
Moreover, the cylinder and helix-based structure factors $G_3$ and $H_2$,
are mathematically identical in form. 
The formal replacement is
\begin{equation}
 M{\bar s} \rightarrow k\Delta z \pm m \Delta \phi,
\end{equation}
and allows us to obtain one from the other.

The extension to double helices is straightforward. 
One merely adds in the potential
for the dyadically related charges and then sums over the $jth$ backbone
charge, $q^j$,
making up the group:
\begin{equation}
\Phi^j_{double-helix}(\rho,\phi,z) =
\Phi^j_{helix}(\rho,\phi+\alpha_j,z+\delta_j)
+\Phi^j_{helix}(\rho,\phi-\alpha_j,z-\delta_j),
\end{equation}
where $2\alpha_j$ and $2\delta_j$ are the offset angles and distances,
respectively \cite{HKE94,EHK94}.

\noindent

{\bf Relation between the helical and cylindrical Green functions}

>From \cite{Jack}, the cylindrical Green function, that is, the 
potential for a unit ($q =1$)
point charge in vacuum ($\epsilon =1$) is given by
\begin{equation}
\frac{1}{|{\bf x} - {\bf x'}|} = \frac{2}{\pi}
\sum_{m= -\infty}^{\infty}\, \int_0^{\infty} dk\,
e^{-im(\phi - \phi')}\cos[k(z - z')] I_m(k \rho_<)K_m(k \rho_>).
\end{equation}
A similar representation for the left hand side of (15) was given
in \cite{LCR} which involves the same set of basis 
functions (essentially products of exponentials and modified
Bessel functions). Thus, a
coordinate transformation should suffice to relate these two
Green functions. 
Since we are talking about the {\em same} 
function, and provided both coordinate representations are
correctly worked out, then 
it must be the case that
\begin{equation}
\frac{1}{|{\bf x} - {\bf x'}|}|_{cylindrical} = 
\frac{1}{|{\bf x} - {\bf x'}|}|_{helical},
\end{equation}
to be understood 
as an equality between one and the same 
function expressed in two
different coordinate systems.
Extending the range in the variable $k$ in the 
above Green function (15) 
by replacing $k \rightarrow |k|$ in the arguments of the Bessel functions, 
use
\begin{equation}
\int_0^{\infty} dk\, \cos[k(z-z')] = 
\frac{1}{2}\int_{-\infty}^{\infty} dk\, e^{-ik(z-z')},
\end{equation}
to rewrite the above Green 
function (15) (see also \cite{Jack}). Now make (or define) the
coordinate transformation $(\phi,z) \rightarrow (t,s)$:
\begin{eqnarray}
z &=& s \sin \beta - t \cos \beta, \\ \nonumber
\phi &=& \frac{s}{a} \cos \beta + \frac{t}{a} \sin \beta,
\end{eqnarray}
which is just the transformation from cylindrical to helical coordinates
defined in \cite{LCR}.
Here, $\beta$ is the pitch angle of the helix, and is related to
the helix pitch by $P = 2\pi a \tan \beta$.
This puts
\begin{equation}
e^{-im(\phi -\phi')} e^{-ik(z -z')} \equiv e^{-iM(s - s')} e^{-iM'(t - t')},
\end{equation}
provided we identify
\begin{eqnarray}
M &=& \frac{m}{a} \cos \beta + k \sin \beta, \\ \nonumber
M' &=& \frac{m}{a} \sin \beta - k \cos \beta.
\end{eqnarray}
This identification can itself be inverted to yield
\begin{eqnarray}
m &=& a(M'\sin \beta + M \cos \beta), \\ \nonumber
k &=& -M'\cos \beta + M \sin \beta.
\end{eqnarray}
In (21), $m$ is an integer while $k$ represents the linear momentum
along the $z$-direction. These are written in terms of $M'$ and $M$, which
represent the components of linear momentum along the helical $t$ 
and $s$-directions, respectively. Using (17), (19) and (21), we can now 
rewrite the helical representation for $\frac{1}{|{\bf x} - {\bf x}'|}$  
calculated in \cite{LCR} 
and transform it back in terms of cylindrical variables. That expression
was given as
\begin{equation}
\frac{1}{|{\bf x} - {\bf x'}|} = \frac{1}{\pi}
\int_{ -\infty}^{\infty}\, dM\, dM'\,
e^{-iM(s - s')}e^{-iM'(t - t')}  I_{\lambda}(k \rho_<)K_{\lambda}(k \rho_>),
\end{equation}
where $\lambda \equiv a(M'\sin \beta + M \cos \beta)$ and $k$ is as above in
(21).
The first step is to insert the identity 
\begin{equation}
\int_{-\infty}^{\infty} dm \, \delta (m - aM'\sin \beta -aM\cos \beta)
=1,
\end{equation}
and perform the integration over $M'$. This gives a constant factor
$(a\sin \beta)^{-1}$; this step is followed by replacing 
the integration over $m$ by a
discrete sum, $\int \frac{dm}{a} \rightarrow \sum_m$, in accord with the
fact that $m/a$ is the component of linear momentum in the $\phi$-direction
of the cylinder. Next, the identity 
\begin{equation}
\int_{-\infty}^{\infty} dk\, \delta(k  - \frac{aM - m\cos \beta}{a \sin \beta})
=1,
\end{equation} 
allows one to carry out the integration over $M$. This produces a 
factor of $\sin \beta$ which cancels the $(\sin \beta)^{-1}$ from the
previous step. Carrying out these steps together with 
the relations in (19) and (20), demonstrates the equivalence (16)
between the two representations calculated for the Green function.

\vfill\eject
\noindent
{\bf Electric field on the central axis} 

Another point of direct comparison with the results of \cite{LCR} is
had by considering the calculation of the
electric field on the central axis of the helix. 
It is not necessary to employ complicated
Green functions
in order to calculate the components of the electric field on the 
central axis in a {\it uniform} medium.
The potential for a single point charge located at ${\bf x}' = (a,\phi',z')$
in a uniform dielectric medium is just
\begin{eqnarray}
\Phi({\bf x}) &=& \frac{q}{\epsilon} \frac{1}{|{\bf x} - {\bf x'}|},\\ 
\nonumber
&=& \frac{q}{\epsilon}
\frac{1}{(\rho^2+a^2-2a\rho \cos(\phi - \phi') + (z - z')^2)^{1/2}}.
\end{eqnarray}
A textbook calculation yields the electric field on the central axis:
\begin{eqnarray}
E_{\rho}(0,\phi,z) &=& -\frac{q}{\epsilon}
\sum_{n,s} \frac{a\cos(\phi - \frac{2\pi s \Delta z}{P})}
{(a^2 + (z - nP - s\Delta z)^2 )^{3/2}}, \\ \nonumber
E_{\phi}(0,\phi,z) &=& \frac{q}{\epsilon}
\sum_{n,s} \frac{a\sin(\phi - \frac{2\pi s \Delta z}{P})}
{(a^2 + (z - nP - s\Delta z)^2 )^{3/2}}, \\ \nonumber
E_z(0,\phi,z) &=& \frac{q}{\epsilon}
\sum_{n,s} \frac{(z - nP - s\Delta z)}
{(a^2 + (z - nP - s\Delta z)^2 )^{3/2}}.
\end{eqnarray}

Since these formulas involve the 
features of the structure of the
DNA (clearly through the two parameters $\Delta z$ and $P$), their
dependence on the lengths and nature of the DNA strands are also obtained
here explicitly. 

The much lengthier calculations of the electric field components on the
(double)-helix central axis in \cite{LCR} lead to the same results only after
that calculation is corrected for a spurious factor of $\sin \beta$ \cite{Ref}.
Indeed, the step invoking the 
replacement of integration over $M'$ in (22) by a sum over $\lambda(\equiv m)$
in \cite{LCR} overlooked this factor of $(\sin \beta)^{-1}$ which
results, as we see, from correct use of the identity (23). Once this is
done, the two sets of electric field components as calculated in terms
of cylindrical and helical coordinates do agree.

Some concluding remarks are in order. First of all, and most importantly 
as demonstrated
above, not only are cylindrical coordinates adequate for calculating
potentials and fields due to helical charge distributions, they also
lead to simple structure factors admitting a straightforward physical
interpretation. 

In regards to the helical coordinate system advocated in \cite{LCR}, it is
a fact that Laplace's equation does not separate in helical coordinates.
Indeed, it has been known for some time that there are exactly eleven 
coordinate systems in which the three-dimensional Laplacian is separable
\cite{MF}; the helical system is not one of them.
Nevertheless, the helical Laplacian in \cite{LCR} was separated assuming
an a-priori, but not general, form for the 
$t$ and $s$-dependent factors of the product ansatz. 
For this reason, it is important to establish the validity of 
the Green function
so constructed and the coordinate transformation between cylindrical and
helical systems leads, as shown above,  
to independent confirmation of the correctness
of the helical Green function. 
At this stage, in so far as one restricts
attention to charged
helices in uniform dielectric media, which coordinate system one prefers to
employ is simply a matter of taste. However, for the more realistic case of a
non-uniform dielectric or solvent, the cylindrical coordinate system offers
an advantage \cite{HKE94,EHK94}. In this situation 
the question of the matching of solutions at the
dielectric interfaces arises. Charges within a 
dielectric induce a surface charge
at the interface between that dielectric and a region with a different
dielectric constant. Since helical coordinates are non-orthogonal and
curvilinear we expect that matching solutions of Laplace's or Poisson's
equations across the boundary will lead to a nontrivial 
albeit solvable problem in
differential geometry. 
We in fact had already considered, but rejected, using helical coordinates
in 1993 for some of the above reasons. 
Finally, even though 
helical coordinates are not separable, we can always take 
the potential and electric field computed in cylindrical
coordinates and then simply transform them to helical coordinates
to find their correct forms in these latter coordinates.

\noindent
{\bf Acknowledgments}

We thank A. K. Rajagopal for a careful reading of the manuscript 
and for making useful comments which helped to clarify 
the connection between the helical and cylindrical Green functions.

\end{document}